\documentclass[aps, prl, twocolumn, notitlepage,superscriptaddress
]{revtex4-2}

\usepackage{graphicx}
\usepackage{dcolumn}
\usepackage{color}
\usepackage{amsmath}
\usepackage{here}
\usepackage{systeme}
\usepackage{xcolor}
\usepackage{multirow}
\usepackage{resizegather}
\usepackage{physunits}
\RequirePackage{natbib}

\begin{document}

\title{Acoustic Analogues of Extremal Rotating Black Holes in Exciton-Polariton Condensates}
\author{Anton Svetlichnyi$^1$, Andrii Chaika$^1$, Alexander Yakimenko}
\affiliation{Department of Physics, Taras Shevchenko National University  of Kyiv,
64/13, Volodymyrska Street, Kyiv 01601, Ukraine} 
\affiliation{Dipartimento di Fisica e Astronomia ’Galileo Galilei’, Universit‘a di Padova, via Marzolo 8, 35131 Padua, Italy}
\affiliation{Istituto Nazionale di Fisica Nucleare, Sezione di Padova, via Marzolo 8, 35131 Padua, Italy }

\begin{abstract}
We theoretically investigate the acoustic analogues of high-angular-momentum rotating black holes in exciton-polariton condensates. Performing numerical simulations of a long-lived ring-shaped condensate configuration with an acoustic horizon and ergoregion for high angular momentum states, we observed a quasi-stable state near critical angular momentum where the acoustic black hole horizon disappears. Our findings offer an insight into the quantum nature of the instability of naked singularity.

\end{abstract}

\maketitle

Analogue gravity provides a unique possibility to explore elusive quantum phenomena near the event horizon or within the ergoregion.  
Numerous approaches have been explored, both theoretically and experimentally, to demonstrate acoustic horizons and analogue of Hawking radiation, including atomic Bose-Einstein condensates (BECs)  \cite{carusotto2008,mayoral2011,lahav2010,steinhauer2016,steinhauer2019,aless2020rampup, PhysRevResearch.2.043065}, ultracold fermions \cite{giovanazzi2005}, superfluid $^{3}$He \cite{jacobson1998}, trapped ions \cite{Horstmann2010}, optical fibers \cite{philbin2008,belgiorno2010,unruh2012,liberati2012}, electromagnetic waveguides\cite{schutzhold2005}, and water tanks \cite{weinfurtner2011,rousseaux2008}.

Among the diverse platforms available for studying analogue gravity, exciton-polariton condensates stand out as especially promising candidates \cite{solnyshkov2011,Jacquet2022,PhysRevLett.114.036402}. These condensates offer an ideal experimental setting to emulate sonic black holes with angular momentum, incorporating key features such as the acoustic event horizon and ergoregion, effectively simulating gravitational properties. The exciton-polariton condensate is a very convenient platform for creating the 2D black analogues \cite{jacquet:hal-02483999, Barcelo:2018ynq}.
Due to the small lifetime of the polaritons, it is easy to establish the required drain of the particles, which is crucial for the creation of a long-lived 2D acoustic black hole. These analogues present a significant opportunity to investigate the intricate quantum properties of black holes, especially those exhibiting high angular momentum states \cite{Jheng:22,Alperin:21} and study their stability in different dissipative and non-dissipative mediums \cite{PhysRevResearch.4.033117,PhysRevA.106.063310}. Moreover, polariton condensates provide the possibility to research the peculiar behavior and configurations of quantum vortices \cite{PhysRevLett.100.250401,Cookson2021}.
 In the recent work \cite{2019PhRvB..99u4511S}, Kerr black holes analogues were theoretically investigated and it revealed some remarkable effects, such as the limitation on the angular momentum of the system, which resembles a restriction for the extremal rotating black holes in general relativity. 
 
 In this work, we investigate the properties of extremal and super extremal sonic black holes in toroidal exciton-polariton condensates. These intriguing systems, suggest an analogue to extremal rotating black holes in general relativity, and give rise to important questions concerning their fundamental properties and the mechanisms governing their eventual decay. 
We demonstrate the stability of acoustic event horizons and ergoregions in a toroidal polariton condensate.
We elucidate the mechanisms governing the decay of extremal black holes and the emergence of analogues to naked singularities in the exciton-polariton condensate, showing the potentially observable signature of quantum vortex dynamics.
We study the impact of non-equilibrium effects on sonic black hole dynamics, shedding light on the puzzling cosmic censorship principle and the possibility of creating and controlling acoustic analogues of rotating black holes.


\begin{figure}[t]
\begin{center}
\includegraphics[width=8.6cm]{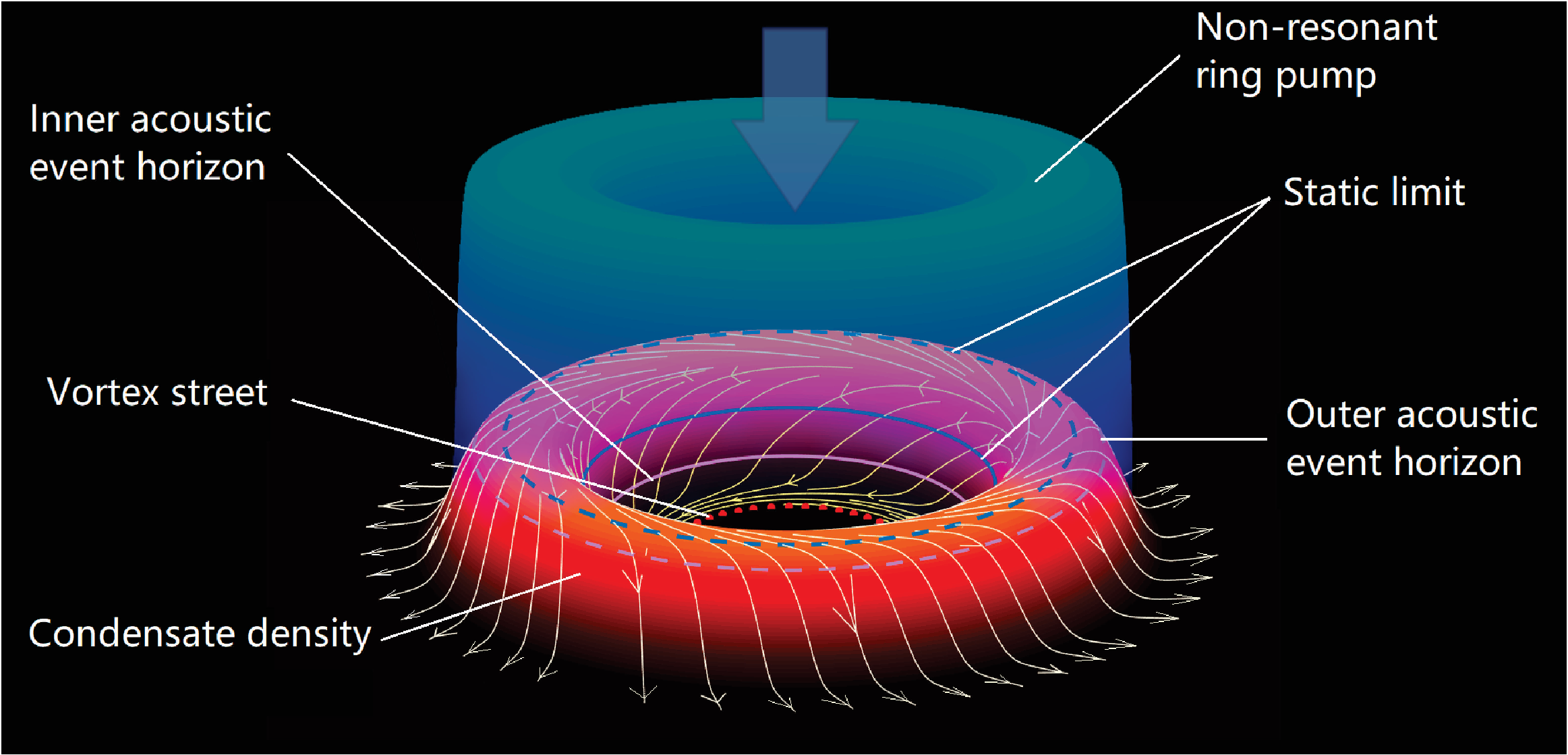}
\end{center}
\vspace{-0.5cm}
\caption{Schematics of the acoustic analogue of rotating black holes in the toroidal exciton-polariton condensate. The blue region is a non-resonant ring pump that creates exciton-polariton condensate (an orange region). The red-doted circle is an $m$-charged vortex street trapped in the central hole of the toroidal condensate. The magenta lines represent the position of the event horizons, whereas bluish circles represent static limits. Solid lines belong to the inner hole, whereas dashed lines to the external hole. The white lines represent the direction of the superflows in the system: from the density peak, they spread inwards and outwards to the internal and external acoustic horizons.}
\label{fig:Schematics3D}
\end{figure}
\emph{\textbf{Model}} -- 
We consider the system schematically shown
in Fig. \ref{fig:Schematics3D}, in which the ring-shaped non-resonant pump creates the condensate with the density gradient on the inner and outer periphery of the ring. Thus, the radial velocity of the condensate can achieve higher values than the speed of sound forming inner and outer acoustic event horizons. The angular momentum in the system is defined by the number of vortices $m$, which can be added by the phase imprinting method, which allows the investigation of the acoustic analogues of rapidly rotating black holes.

The non-resonantly pumped polariton condensate can be described by the mean-field dissipative GPE for the macroscopic wavefunction, $\psi$, coupled to the rate equation for the density of the excitonic reservoir, $n_R$ (see, e.g. \cite{Wouters_sos,PhysRevB.91.184518}):

\begin{align} \label{eq:GPE}
i \hbar\frac{\partial \psi}{\partial t}=\big[-\frac{\hbar^2}{2M_0}\nabla^2+g_c |\psi|^2+g_R n_R \nonumber\\
\qquad\qquad\qquad+\frac{i \hbar}{2}\left( R\, n_R-\gamma_c \right) \big] \psi,\\
\frac{\partial n_R}{\partial t}=P(\mathbf{r},t)-\left(\gamma_R+R |\psi|^2\right) n_R, \nonumber 
\end{align}
where $P(\textbf{r},t)$ is the optical pumping rate, $g_c$ and $g_R$ characterise polariton-polariton and polariton-exciton interactions, respectively. The relaxation rates $\gamma_c$ and $\gamma_R$ quantify the finite lifetime of condensed polaritons and the reservoir, respectively. The stimulated scattering rate,  $R$, controls the growth of the condensate density. Note that in comparison with the model employed in Ref. \cite{2019PhRvB..99u4511S} for analysis of a similar system, our approach involves coupled equations that capture the dynamic evolution of both the condensate wave function and the excitonic reservoir. To achieve a long-lived state with significant angular momentum, we explore various shapes of toroidal pumping $P(\textbf{r},t)$.

\begin{figure}[h!]
    \centering
    \includegraphics[width=8.6cm]{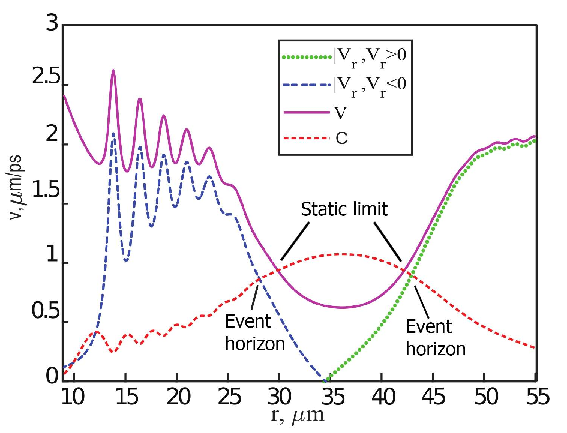}
    \caption{The radial distribution of the full velocity (magenta line) $v$ and radial component of the velocity (blue line for $v_{r} < 0$ and green line for $v_{r} > 0$) of the condensate along with the speed of sound (red line) $c$. The intersection points of the speed of sound with full and radial velocity are the static limits and acoustic horizons of the black holes respectively.}
    \label{fig:schem}
\end{figure}

\emph{\textbf{Long-lived configuration of the black hole}} -- 
To explore the analogue of a rotating black hole, we establish an acoustic analogue of the static limit and the event horizon. The static limit arises when the Bogoliubov speed of sound, given by \(c = \sqrt{{g_c|\Psi|^2}/{M_0}}\), matches the full velocity of the condensate, denoted as \(\textbf{v} = (\hbar/M_0)\nabla \text{arg}(\Psi)\). The existence of an acoustic horizon occurs in the region where \(c = v_{r}\), with \(v_r\) representing the radial velocity of the condensate. Creating a long-lived analogue of a black hole in polariton condensates requires a radially symmetric pump that effectively drains polaritons in the near-center region of the toroidal condensate. The essential properties of sonic black holes are governed by the condensate density gradient which can be tuned by the shape of the pump. In this work, we employed two types of pumping to achieve long-lived states with high angular momentum.
First, we consider a simple ring-shaped pump:
 \begin{equation} \label{rigid_pump}
    P = P_{0}\; \exp\left\{-\left(\frac{r-r_0}{w}\right)^{4}\right\},
\end{equation}
where $P_{0} = 65\, \meter[\micro]^{-1} {\s[p]}^{-1}$ - is the optical pumping amplitude, $r_0 = 36 \, \meter[\micro]$,  $w = 10 \, \meter[\micro]$ - controls the width of the ring. First of all, we created the stable non-extremal rotating black hole with $m = 19$, to study the properties of this peculiar system. Due to the ring topology of the pump, there are two acoustic horizons and static limits in the system: on the outer and inner periphery of the ring [Fig. \ref{fig:schem}]. But the distance between the outer static limit and the horizon is too small ($\approx 1 \meter[\micro]$) so it becomes hard to study the properties of the ergosphere in that case. That is why we concentrate our studies on the black hole on the inner periphery, where the analogy to the real black holes can be explored more precisely.

In our model, one is free to choose the initial conditions, due to the short lifetime of the polaritons, but for convenience, we decided to start with random noise to accelerate the creation of a stable configuration.  To observe the long-live solution as in [Fig. \ref{fig:schem}.(b)], we wait for 40-50 ps for currents to stabilize. The lifetime of such systems is usually up to 1000 ps. \\
\begin{figure}[h!]
    \centering
    \includegraphics[width=8.6cm]{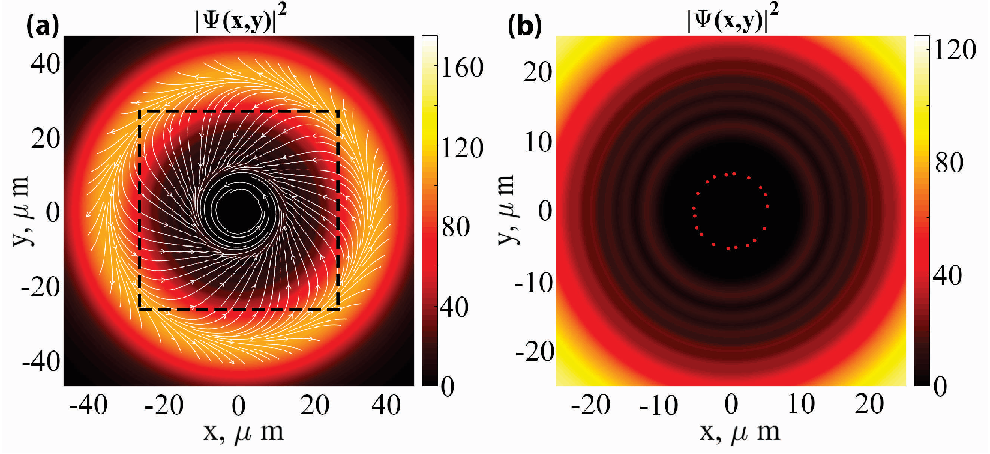}
    \caption{(a) Density of states for the stable configuration ($m = 19$) along with contour lines of the total velocity (white lines) of the condensate. (b) Bessel modes inside the highlighted region of \ref{fig:bes}(a) region. Red points denote vortices. }
    \label{fig:bes}
\end{figure}
\emph{\textbf{Bessel modes}} -- 
There are several very peculiar properties in that system. First is the standing wave (set of density circles) inside the black hole region [Fig. \ref{fig:bes}(b)], similar to observed in \cite{Ma2018_Bessel_modes}. This effect is connected with polaritons creation and decay in the low-density region and it is solely the property of the exciton-polariton condensate. This effect can be explained in terms of the low-density approximation.

We are searching for stationary solutions of the system in the form of standard ansatz:
\begin{equation} \label{eq: Stationary solution}
 \begin{aligned}
 \psi(r,t) = \Psi(r) e^{- i\mu t / \hbar}, \\n_{R}(r)=\frac{P(r)}{\gamma_r + R|\Psi|^2}.
 \end{aligned}
\end{equation}
Due to the absence of a pump in the decaying region, we can neglect pumping and non-linear terms in Eq. (\ref{eq:GPE}). This approximation gives us:
\begin{equation} \label{Bessel modes 1}
    \mu \Psi= \left(-\frac{\hbar^2}{2M_0}\nabla^2 -\frac{i \hbar}{2}\gamma_c \right) \Psi. 
\end{equation}
This is the well-known Helmholtz equation. In the central decaying region solution which can be written in terms of Bessel functions of the first kind:
\begin{equation} \label{Bessel modes 2}
    \Psi(r) \sim J_m\left(kr\right) e^{i m\phi}.
\end{equation}
Consequently, for the external region, we can write the solution in the form of the Hankel functions of the first kind:
\begin{equation} \label{Bessel modes 3}
     \Psi(r) \sim H_m^1(k r) \ e^{i m\phi} \approx \frac{e^{i(k r + m \phi)}}{\sqrt{k r}},
\end{equation}
where $k=\sqrt{\widetilde{\mu}+iM_0 \gamma_c/\hbar}$ and $\widetilde{\mu}=2 M_0\mu/\hbar^2 $.
Here the wave number exhibits a complex argument, however, in our simulations, the imaginary part is usually 15-30 times smaller than the absolute value. Consequently, within the low-density region, the wave function demonstrates behaviour closely resembling that of a Bessel function with a real argument (see Fig. \ref{fig:bes}).

\begin{figure}[h!]
    \centering
    \includegraphics[width=8.6cm]{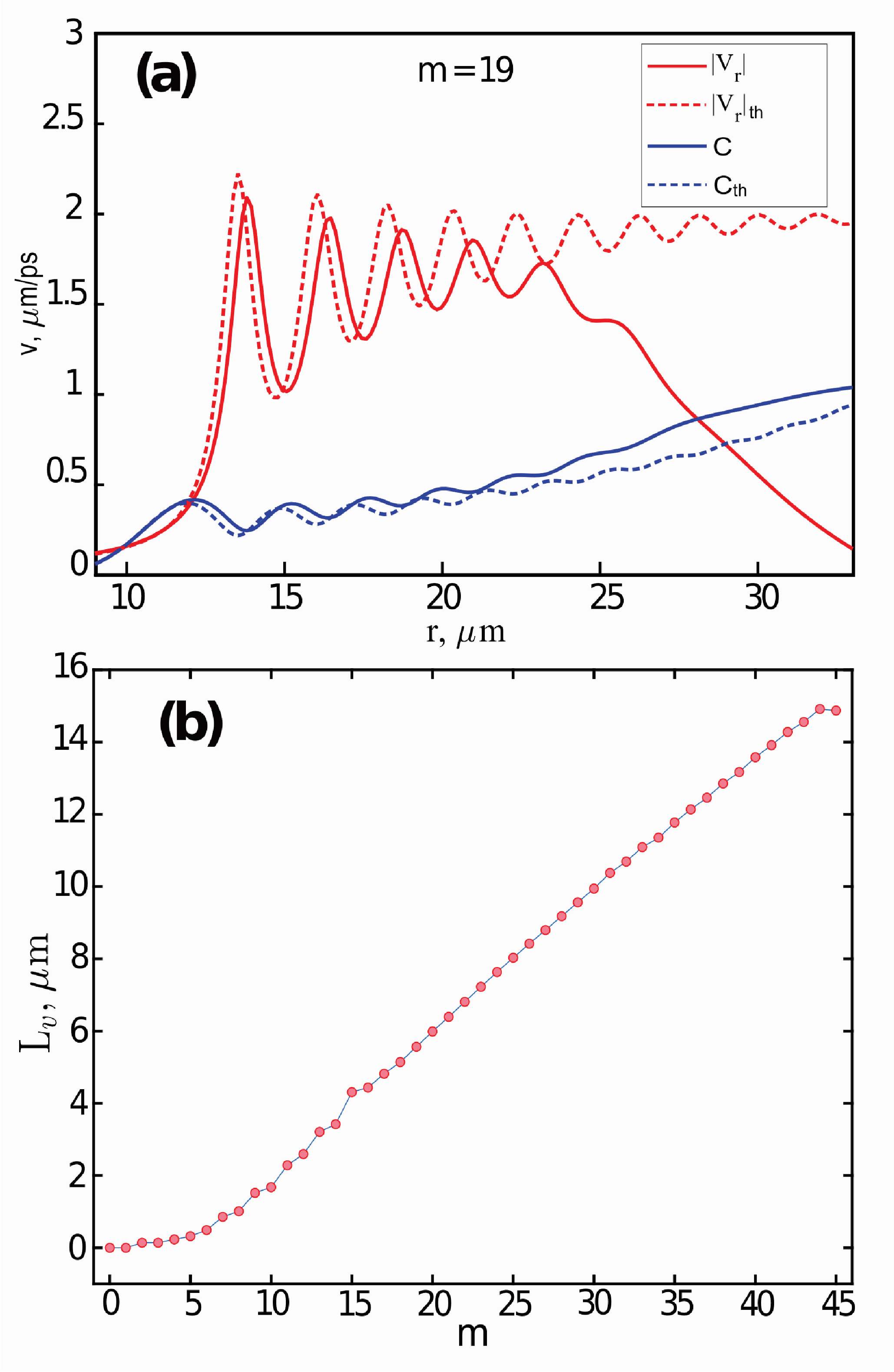}
    \caption{(a) The comparison of the approximate estimate for the speed of sound (dashed blue line) and radial velocity (dashed red line) with the results of the numerical simulations (solid red and blue lines, correspondingly). (b) The vortex street radius $L$ vs the topological charge $m$ for the pumping shape given by Eq.  (\ref{rigid_pump}).}
    \label{fig:compr}
    \end{figure}

It is worth mentioning that the imaginary part of $k$ yields the non-zero radial velocity. The comparison between theoretical estimation and results of the simulations is presented in Fig. \ref{fig:compr}(a). One can see a good agreement between the radial velocity of the condensate and the speed of sound inside the low-density region. The divergences are caused by the influence of the pumping and non-linear effects.\\

\emph{\textbf{Critical angular momentum of the black hole}} -- 
One of the most remarkable features of the ring-shaped exciton-polariton condensates is the formation of the vortex street observed in experiments \cite{Boulier2015} [Fig. \ref{fig:bes}(b)]. 
The radius of the vortex street increases linearly with topological charge  [see Fig. \ref{fig:compr}(b)], which leads to the decay of the persistent current above some critical value of the angular momentum.

 As mentioned above, we concentrate our studies on the inner black hole region with the flow directed to the center of the ring, i.e. with \emph{negative} radial velocity. 
 From Eq. (\ref{eq:GPE}) one can obtain the following continuity equation for the stationary state Eq. \eqref{eq: Stationary solution}:
 \begin{equation} \label{continuity equation 0}
    \nabla \left(\rho \vec{v} \right) = \rho \left(-\gamma_c + \frac{PR}{\gamma_R+R \rho} \right).
\end{equation}
At the right side of \eqref{continuity equation 0} are the "source" and "sink" terms of the condensate. We call the line at which the right side vanishes the gain-loss border. Also for convenience, we call the line at which radial velocity changes the sign ($v_r=0$) "the watershed". Integrating \eqref{continuity equation 0} over the internal region with negative radial velocity, up to "the watershed" and applying the divergence theorem we get:

\begin{equation} \label{continuity equation 2}
   \int \rho  \left(-\gamma_c + \frac{PR}{\gamma_R+R \rho} \right) d S = 0.
\end{equation}
That expression describes the balance of the flow generation in the sink and source regions.
When the system gains additional angular momentum the condensate is forced out from the center by centrifugal force, and the density in the sink region is decreasing faster, than in the source region. To fulfill the balance, in \eqref{continuity equation 2}, the area of the source part must decrease, so the watershed shifts closer to the axis of the ring.  \\ %

At high angular momentum due to the proximity to the vortex street the density near the gain-loss border is negligible, and the radial coordinate $r_*$ of the critical gain-loss border can be obtained from the following equation:
\begin{equation} \label{continuity equation 3}
 P(r_*) = \frac{\gamma_c \gamma_R}{R}. 
\end{equation}

\begin{figure}[h!]
    \centering
    \includegraphics[width=8.6cm]{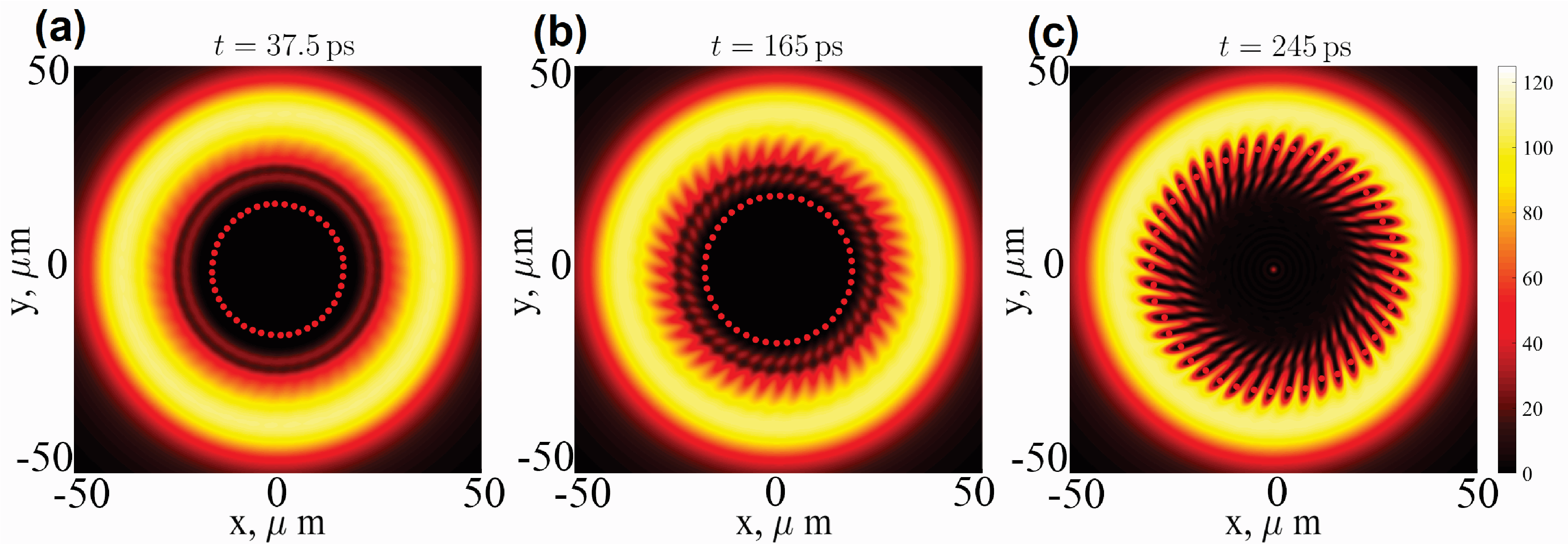}
    \caption{Density $|\psi|^2$ for the pump (\ref{rigid_pump}) illustrating unstable system evolution with $m=44$ at different times. The red points denote vortex core positions.}
    \label{fig:decay_density}
\end{figure}

Our simulations reveal that when the vortex street approaches the gain-loss boundary, the radial velocity of the condensate turns unidirectional, leading to the disappearance of the inner black hole's acoustic horizon -- a behavior analogous to a naked singularity in general relativity. However, in practice, vortices exiting the condensate typically prevent this state, as illustrated in Fig. \ref{fig:decay_density}. Specifically, with the pump (\ref{rigid_pump}), our simulations demonstrate a maximum angular momentum of $m = 43$ and a vortex street radius of $L = 15.8\, \meter[\micro]$, which is lesser than the critical gain-loss border, $r_* = 25.2\, \meter[\micro]$, hence vortices start leaving the condensate before the event horizon disappears.

\begin{figure}[h!]
    \centering
    \includegraphics[width=8.6cm]{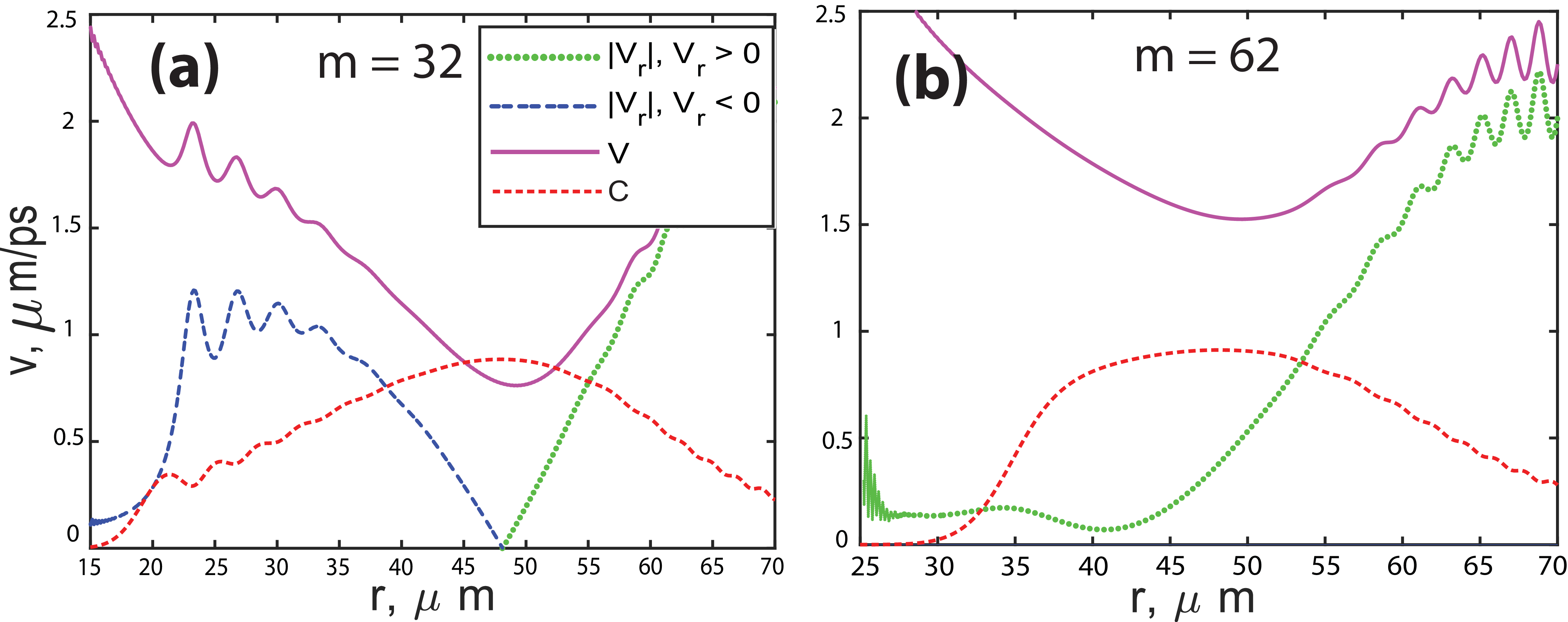}
    \caption{The velocity of the condensate and speed of sound dynamics for pump (\ref{Mapam}): (a) m = 32, long-lived state. (b) m = 62, the state with unidirectional flows - complete disappearance of the inner event horizon.}
    \label{fig:vel_ma}
\end{figure}
\begin{figure}[h!]
    \centering
    \vspace{-0.5cm}
    \includegraphics[scale = 0.34]{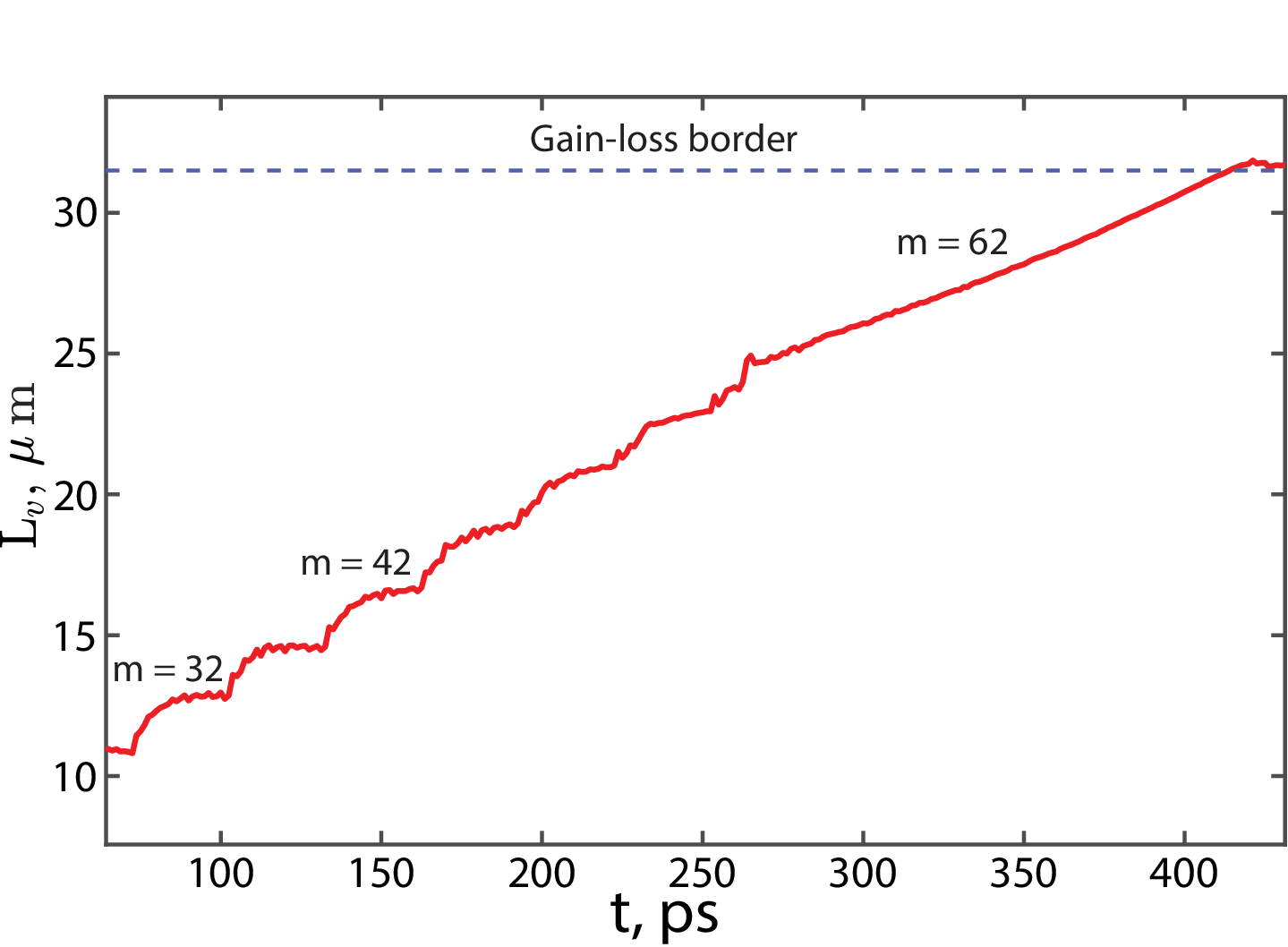}
    \caption{The time-dependent radius of the vortex street (red line) and the critical gain-loss border (blue dashed line). We imprint five vortices every $30 \s[p]$  until radial flows become unidirectional ($m = 62$). Upon reaching the gain-loss boundary, the ring-shaped vortex street decays, and vortex lines escape the condensate.}
    \label{fig:rvr}
\end{figure}

 \emph{\textbf{High angular momentum state}} -- 
 In our studies, we tested several pumps to create a black hole analogue. To demonstrate the disappearance of the horizon above some threshold angular momentum, we introduced the following ring-shaped pump, which is similar to the stable pump, studied in \cite{Mapump}:
\begin{equation}\label{Mapam}
    P = P_{0}\;\left(1 - \exp\left\{-\left(\dfrac{r}{w}\right)^{4} \right\}\right)\exp\left\{-\left(\dfrac{r}{w}\right)^{8}\right\},
\end{equation}
where $P_{0} = 170 \, \meter[\micro]^{-1} {\s[p]}^{-1}$ - is the optical pumping amplitude, $w = 56 \, \meter[\micro]$ is the width of the ring. At low momentum ($m = 32$), we observe a robust acoustic analogue of a rotating black hole [Fig. \ref{fig:vel_ma}(a)]. To further enhance the system's angular momentum, we systematically introduce five vortices at the center of the ring with a time interval of $30$ ps [Fig. \ref{fig:rvr}].
\begin{figure}[h!]
   \centering
   \includegraphics[width=8.6cm]{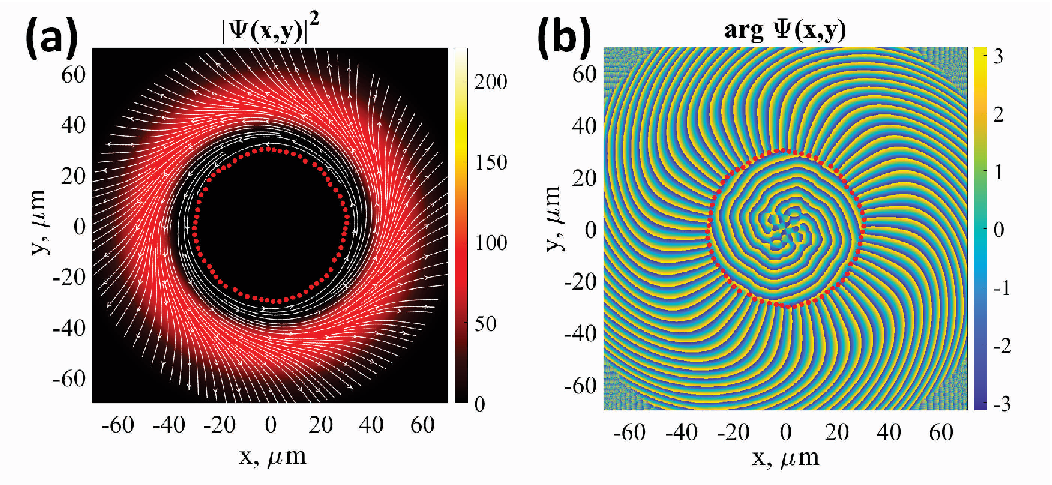}
   \vspace{-0.5cm}
    \caption{(a) Density $|\Psi|^2$ with overlaid streamlines representing the condensate flow (white lines with arrows), notably transitioning into a unidirectional state. (b) Condensate phase, highlighting the ring-shaped vortex street. Positions of the vortex cores are indicated by red points.}
    \label{fig:ns}
\end{figure}

As noted, as the ring-shaped vortex street expands toward the horizon, the internal region between the horizon and vortex cores diminishes [Fig. \ref{fig:vel_ma}(b)]. A noteworthy phenomenon occurs when the vortex street meets the acoustic horizon: the black hole horizon vanishes, and the radial flow along the to-axis reverses its direction, resulting in a unidirectional out-axis flow [Fig. \ref{fig:ns}(a)].
The ring-shaped vortex street can be analogously interpreted as a representation of a ring-shaped singularity in the Kerr black hole. Therefore this state with unidirectional flow can be treated as an analogue of a super-extremal black hole with a naked singularity.



In our simulations, we observed that state with the vortex street radius less than the critical gain-loss border $r_* = 31.5\, \meter[\micro]$ due to the finite size of vortices, which is not taken into account in Eq. (\ref{continuity equation 2}). The divergence is an order of the characteristic length between the vortices $\approx 2.5\, \meter[\micro]$.


It is important to highlight that the current state of the system is metastable, and characterized by a limited lifetime. As illustrated in Fig. \ref{fig:rvr}, the radius of the vortex street grows throughout $100  \s[p]$. Subsequently, beyond this timeframe, vortices begin leaving the condensate.
 Note that the number of vortices in the vortex street on Fig. \ref{fig:ns} is equal to 68, not 62 as the total topological charge of the system. Inside the low-density region, there is a constant process of vortex-antivortex pairs annihilation and creation process. Due to the noisy phase in the center of the system, we decided not to depict these vortices and antivortices in Fig. \ref{fig:ns}.  The extra six vortices in the vortex street are connected with the corresponding antivortices inside the central low-density region of the toroidal condensate, so the full topological charge is conserved. 


\emph{\textbf{Conclusions}} -- 
We investigated acoustic analogues of rotating black hole event horizons formed by the persistent currents in the ring-shaped exciton-polariton condensate with two acoustic event horizons formed by the radial superflows of the condensate.  
We have found that above a critical angular momentum threshold, a reversal in radial flow occurred. Intriguingly, this transition coincided with the disappearance of the internal acoustic event horizon resembling a hypothetical super-extremal black hole (see, e.g. \cite{PhysRevD.84.064044}). However, such a scenario is unlikely to occur in nature, as super-extremal black holes are unstable and would quickly lose angular momentum through radiation or accretion. Theoretically predicted in our work long-lived transitory state can be treated as  
an acoustic analogue of a black hole with \emph{naked singularity}. These unusual states finally decay so that the 'cosmic censorship principle' appears to be fulfilled for the stable stationary states.


Our findings hold promise for a profound exploration of the quantum properties of Kerr black holes using acoustic analogues within well-controlled laboratory environments of exciton-polariton condensates.

We acknowledge Luca Salascnich and Elena Ostrovskaya for the useful discussions. A Y and A S
acknowledge support from the National Research Foundation of Ukraine through Grant No. 2020.02/0032.  A Y acknowledges support from the BIRD Project ‘Ultracold atoms in curved geometries’ of the
University of Padova.

\bibliography{references_EBH}

\end{document}